# Productivity within the ETAS seismicity model

G. Molchan[1], E. Varini[2], A. Peresan[3]

[1] Institute of Earthquake Prediction Theory and Mathematical Geophysics,
Russian Academy of Science, 84/32 Profsoyuznaya st.,117997, Moscow, RF.
Email molchan@mitp.ru

[2] Institute for Applied Mathematics and Information Technologies "Enrico Magenes",
National Research Council, via Corti 12, 20133 Milano, IT, Email elisa.varini@cnr.it

[3] Seismological Research Centre, National Institute of Oceanography and Applied
Geophysics, via Treviso 55, 33100, Udine, IT, Email aperesan@inogs.it

*26 April 2022*

**SUMMARY**

The productivity of a magnitude $m$ event can be characterized in term of triggered events of magnitude above $m-\Delta$: it is the number of direct 'descendants' $\nu_\Delta$ and the number of all 'descendants' $V_\Delta$. There is evidence in favor of the discrete exponential distribution for both $\nu_\Delta$ and $V_\Delta$ with a dominant initial magnitude $m$ (the case of aftershock cluster). We consider the general Epidemic Type Aftershock Sequence (ETAS) model adapted to any distribution of $\nu_\Delta$. It turns out that the branching structure of the model excludes the possibility of having exponential distributions for both productivity characteristics at once. We have analytically investigated the features of the $V_\Delta$ distribution within a wide class of ETAS models. We show the fundamental difference in tail behavior of the $V_\Delta$-distributions for general-type clusters and clusters with a dominant initial magnitude: the tail is heavy in the former case and light in the latter. The real data demonstrate the possibilities of this kind. This result provides theoretical and practical constraints for distributional analysis of $V_\Delta$.

## 1 INTRODUCTION

Earthquakes clustering, aftershocks occurrence in particular, is a prominent feature in observed seismicity. Different models have been proposed to describe the features of triggered events, especially in the case of aftershocks clusters with an initial dominant magnitude earthquake; yet, several theoretical aspects still need to be explored. In this study we investigate the distributions of the number of triggered events as described by a branching process, which might provide useful practical constraints for empirical analysis of earthquake data.

Let $V_\Delta$ denote the number of seismic events of magnitude $m \geq m_0 - \Delta$ caused by an earthquake of magnitude $m_0 \geq m_c + \Delta$, where $m_c$ is the lower magnitude threshold and $\Delta$ a positive default value. $V_\Delta$ is further referred to as the *total $\Delta$-productivity*.

Considering an earthquake cluster as the result of a branching process [Zaliapin et al., 2008], we can also deal with the directly triggered events of $m_0$, i.e. the first generation events of an $m_0$-event. The number $v_\Delta$ of such events is in the same magnitude range $m \geq m_0 - \Delta$ and is later named $\Delta$-*productivity* of $m_0$.

The distributions of quantities $V_\Delta$ and $v_\Delta$ have recently become the object of active analysis in the literature (Baranov and Shebalin, 2019; Shebalin et al., 2020; Shebalin and Baranov, 2021; Baranov et al., 2022). According to these publications the $v_\Delta$-distribution $f(n) = P(v_\Delta = n)$ is *universal in shape*, namely, it is *geometric* (or *discrete exponential*):

$$f_G(n) = p^n(1-p) \qquad (1)$$

where $p$ does not depend on the magnitude of triggering event. This result is relevant in statistical seismology because the popular Epidemic Type Aftershock sequence (ETAS) model (Ogata, 1998) is based on the Poisson distribution

$$f_P(n) = e^{-\lambda} \lambda^n / n! \qquad (2)$$

an assumption that has never been tested.

Similarly, the geometric distribution (1) was adopted for statistics $V_\Delta$, provided that it refers to clusters with a dominant initial magnitude $m_0$, that is, to aftershocks. For the first time this hypothesis appeared in the work by Solovyev and Solovyeva [1962]; it was based on relatively scarce data (1954-1961) for the Pacific Belt ($m_o \approx 7$; $\Delta = 2$) and for the Kamchatka and Kuril Islands region ($m_o \approx 6$; $\Delta = 2$). Shebalin et al. [2018] expanded the analysis and confirmed the hypothesis using 850 aftershock sequences with main shocks $m_o \geq 6.5$, $\Delta = 2$ from the ANSS (1975-2018) catalog. In the same time, Kagan [2010], using aftershocks $m \geq 4.7$ from PDE (1977-2007) catalog for $m_o = 7.1 - 7.2$, found that the empirical distribution of $V_\Delta$ is bimodal with the dominant peak at $n = 0$.

Zaliapin and Ben-Zion (2016) provided a thorough analysis of earthquake clustering and showed its spatial dependence and connection with heat flow production. They showed that, on the global scale, the total $\Delta$-productivity distribution (for clusters with $\max m \geq 6$ and $\Delta = 2$) looks like a power-law for locations with high heat flow $(H > 0.2W/m^2)$, and as a log-normal distribution otherwise. These conclusions are qualitative because they are based on log-log plots. Nevertheless, the data for the regions of high heat flow differ significantly from the geometric hypothesis (1) for aftershocks. The same should be expected for any cluster with max $m \leq 5$. In this case, according to Zaliapin and Ben-Zion (2016), the cluster size distribution in zones with both high and low heat flow is a power type distribution, $f(n) \propto n^{-\kappa-1}, \kappa \approx 2.5$.

These observations rise the following questions: a) is it possible to consider the geometric distributions of $\nu_\Delta$ and $V_\Delta$ (for aftershocks) as laws of seismicity; and b) how to explain the difference in the tail behavior of the $V_\Delta$ distribution for arbitrary clusters and for aftershocks. A rigorous analysis of these issues is possible within the framework of a model adequately related to the concepts of $\nu_\Delta$ and $V_\Delta$. A model suitable for this purpose may be the ETAS model with an arbitrary (not only Poisson) distribution of direct aftershocks associated to any event (Saichev et al, 2005). The purpose of this note is twofold: first, to show the incompatibility of the hypotheses about the geometric distribution of $\nu_\Delta$ and $V_\Delta$ (for aftershocks) within the general ETAS model; and second, to prove the fundamental difference between size distributions for general type clusters and aftershock clusters.

## *2 ETAS MODELS*

Let us consider the seismic events $x = (t, g, m)$ occurred in the time-location-magnitude space $S = (t \geq 0, g \in R^2, m \geq 0)$. The lower magnitude threshold $m_c$ is taken here as the zero reference point without loss of generality. According to the branching structure of the ETAS model, an earthquake cluster can be represented as follows (Saichev et al, 2005). The initial event $x_0 = (t_0, g_0, m_0)$ generates a random number $\nu(m_0)$ of events $\{x_i, i = 1,..., \nu(m_0)\}$ where $\nu(m_0)$ follows the special distribution F and the component $\lambda(m_0) = E\nu(m_0)$. The offspring events are distributed in $S$ as independent variables with the probability density:

$$f(x|x_0) = f_1(m) f_2(t - t_0) f_3(g - g_0 | m_0). \tag{3}$$

Each new event independently generates its own family of events following the generation law described above, and so on. If the mean of the direct triggered events is $\Lambda_1 = E\lambda(m) < 1$, then the resulting cluster is almost surely finite.

The independence of the magnitudes of direct 'descendants' between themselves and from the parent event $x_0$ is an important property of the model. This allows us to consider the

Productivity within the ETAS seismicity model distribution $f_1(m)$ as a frequency-magnitude law for the cluster as a whole. This also is important for understanding the choice of the distribution models for $v_\Delta$ and $V_\Delta$.

The components of the *regular ETAS model* are specified as follows:

$v(m)$ -distribution — Poisson

$\lambda(m) = \lambda e^{\alpha \cdot m}, m \geq 0$ — a copy of the Utsu law for aftershocks

$f_1(m) = \beta \exp(-\beta m)$, $m \geq 0$; $\beta > \alpha$ — the Gutenberg-Richter law

$f_2(t) = (p-1)(t/c+1)^{-p}/c$ — the time scattering (a copy of the Omori law for aftershocks)

$f_3(g|m) = (q-1)[|g/\psi_m|^2 + 1]^{-q}/(\pi \psi_m^2)$ — one of the laws of scattering in space,

$\psi_m = d e^{\gamma \cdot m/2}$ — the scale parameter.

*Note*: The exponents $(\alpha, \beta)$ in the laws of Utsu and Gutenberg-Richter are more convenient (Saichev et al, 2005) to use with the base "e" instead of the usual 10.

A theoretical analysis of various aspects of the general ETAS model with the regular components $\lambda(m)$ and $f_1(m)$ is provided in the works of Saichev and co-authors (see Saichev and Sornette, 2017 for further details). In particular, they proved the power law behavior of the cluster size distribution $f_{cl}(n) \propto n^{-1-\kappa}$ with $\kappa \in [0.5, 1]$ (Saichev et al, 2005; Baró, 2020)). Note that, according to Zaliapin and Ben-Zion (2016), $\kappa \in [1, 2.5]$ for real data.

In the general case we need to specify the family of $v(m)$ distributions: $F = \{f(n|m)\}$. This choice will be reflected in the designation of the model as ETAS(F). In particular, we will set F=P for Poisson, and F=G for Geometric distribution. The characteristic $\lambda(m) = Ev(m)$ will be common to all models. The relation of $\lambda(m)$ with the parameter $p(m)$ of the geometric distribution is determined by the equality $\lambda = p/(1-p)$, i.e.

$$p(m) = \lambda(m)/(1+\lambda(m)) \qquad (4)$$

Recall that the regular ETAS model can be defined in terms of the conditional intensity of events (Hawkes, Oakes, 1974)

$$\lambda^*(x|\{x_i : t_i < t\}) = \mu(g) f_1(m) + \sum_i \lambda(m_i) f(x|x_i), \qquad (5)$$

where $x = (t, g, m)$ and $\mu(g)$ is the rate of background seismicity. The above definition is broader because it allows for any $v(m)$ distribution.

To provide the G-distribution of $\Delta$-productivity for any $\Delta$, it is natural to consider the G-distribution of $v(m)$. It follows from the following seemingly obvious but useful statement.

G. Molchan, E. Varini, A. Peresan

**Statement 1**. *Let the distribution type F of $\nu(m_0)$ is F=G or P. Then $\nu_\Delta(m)$-distribution will have the same type for $m_0 \geq \Delta$. In both cases the statistics $\nu_\Delta(m_0)$ and $\nu(m_0)$ differ only in average values:*

$$\lambda_\Delta(m_0) = \lambda(m_0) \overline{F_1}(m_0 - \Delta) / \overline{F_1}(\Delta) \qquad (6)$$

where $\overline{F_1}(m) = \int_m^\infty f_1(u) du$, $\overline{F_1}(0) = 1$.

Let $a = f_1(0) > 0$ and $\lambda_\Delta(m_0)$ be independent of $m_0$ in the area $0 < \Delta \leq m_0 < M$. Then $\lambda(m) = ce^{am}$ and $\overline{F_1}(m) = e^{-am}$, $m \in (0, M)$.

The proof of the statement and all subsequent ones are contained in the Appendix.

**Remark 1.** The conditions for the independence of the $\nu_\Delta(m_0)$-distribution from $m_0$ are very strict, and therefore, in contrast with Shebalin et al. (2020), this property should be unstable.

**Remark 2: RT property**. Statement 1 allows the following generalization. Consider the $RT(\pi)$ operation of a *random cluster thinning*: $RT(\pi)$ saves each cluster event independently with probability $\pi$, and deletes it with probability $1 - \pi$; the transformed cluster size $V$ is $RT(\pi)V$. Given the magnitude independence in the ETAS(F) model, the productivity $\nu_\Delta$ has the same distribution as $RT(\pi)\nu$ provided that $\pi = P(m > m_0 - \Delta)$.

In the case F=(P or G), $E[RT(\pi)\nu] = \pi E\nu$ and the distribution *type* of $\nu$ and $RT(\pi)\nu$ is identical. We will call these two properties of the discrete distribution the *RT property*. We can formalize the *RT*- property as follows.

Let $V$ and $V(\pi) = RT(\pi)V$ be the cluster sizes, and assume that $\lambda = EV$ and $Ez^V = \varphi(z, \lambda)$. Then by (A1) from Appendix, $Ez^{V(\pi)} = \varphi(1 - \pi + \pi z, \lambda), EV(\pi) = \lambda\pi$. The RT property implies that

$$\varphi(1 - \pi + z\pi, \lambda) = \varphi(z, \lambda\pi).$$

Setting $\lambda\pi = a$ and $(w-1)/a = (z-1)/\lambda$, we get the following criterion of the RT property:

$$\varphi(w, \lambda) = \psi(\lambda(w-1)) \qquad (7)$$

where $\psi(z) = \varphi(1 + z/a, a)$ is analytic in the disk $|z + a| < a$ for any $\lambda > a$, and is continuous at $z = 0$. Any such function $\psi(z)$ specifies the type of F in the ETAS(F) model for which statement 2 remains true.

Productivity within the ETAS seismicity model

*Examples of the RT property*

1) The Negative Binomial Distribution (NBD) for which $\varphi(z,\lambda) = (1-\lambda(z-1))^{-\tau}, \tau > 0$ holds the RT property. In the case $\tau = 1$ it corresponds to the Geometric distribution.

*2) Combining N independent clusters with RT property also preserves this property.* Formally this follows from (7) and the relation for the cluster characteristics $\varphi_i(z,\lambda_i)$:

$$\varphi(z,\lambda) = \prod_i \varphi_i(z,\lambda_i) = \prod_i \psi_i(p_i\lambda(z-1)), \quad p_i\lambda = \lambda_i \text{ and } \sum p_i = 1. \tag{8}$$

The NBD model was used by Kagan (2014) to describe the distribution of annual earthquake numbers for the CIT catalogue 1932-2001, m>5.0. Below we will show that the RT-property, which NBD possesses, is poorly compatible with the branching cluster structure. However, in the Kagan's case, the statistics of events occurring worldwide in one year consists mainly of independent weakly grouped events. Such case is closer to the model (8) with components: $\varphi_i(z,\lambda_i) = 1 - \lambda_i(z-1), \lambda_i < 1$.

3) *A cluster selected randomly with probabilities $\{\rho_i > 0\}$ from N clusters with RT property retains RT property*. In this case

$$\varphi(z,\lambda) = \sum_i \varphi_i(z,\lambda_i)\rho_i = \sum_i \psi_i(p_i\lambda(z-1))\rho_i, \quad \lambda p_i = \lambda_i \text{ and } \sum \lambda_i \rho_i = \lambda. \tag{9}$$

However*, if the cluster sizes distributions are of the same type*, that is. $\varphi_i(z,\lambda_i) = \psi(\lambda_i(z-1))$, *then the random cluster usually loses this type provided that $\lambda_i \neq const$*.

For example, assuming a geometric distribution for both the size of a random cluster and the sizes of its components, (9) gives $(1-w)^{-1} = \sum_i (1-wp_i)^{-1}\rho_i, w = \lambda(z-1)$, which is possible only when $p_i = 1$. This is also true in general case if $\varphi(z,\lambda_i)$ is N-smooth at z=1

The considered case is typical for the observed clusters taken as one statistical population. Consequently, empirical distributions of $v_\Delta$ or $V_\Delta$ obtained over large territories generally do not represent the true types of their distributions. This circumstance makes it difficult to prove the universality of the type of $v_\Delta(m_0)$ distribution.

**Remark 3: ETAS\* model.** To get the geometric distribution of $v_\Delta$ in framework of the ETAS model, Shebalin et al. (2020) used randomization of the parameter $\lambda(m_0)$. They used the fact that a Poisson distribution with an exponentially distributed random parameter is equivalent to a geometric distribution. The resulting model was named ETAS\*. Its analogue is the ETAS(G) model in which the average productivity does not depend on the initial magnitude, namely $\lambda(m_0) = \Lambda$. Because of (6), in this case $\lambda_\Delta(m_0)$ becomes dependent on $m_0$ for any $f_1(m) > 0$.

G. Molchan, E. Varini, A. Peresan

The analogue of the ETAS* model may be interesting due to its connection with the classical branching Galton-Watson model (Harris, 1963). Both models are identical if we omit the space-time component in the model. This gives us an exact formula for the distribution of the total productivity of any initial event (cluster size not including the initial event):

$$P(V = n) = \Gamma(1/2 + n)/\Gamma(2 + n) \times (4pq)^n (q/\sqrt{\pi}), \quad q = 1 - p \qquad (10)$$

where $\Gamma(x)$ is the Gamma function, and $p < 1/2$ is the parameter of the geometric distribution (Harris, 1963). Formula (10) remains valid in the critical regime when $\lambda = p/(1-p) = 1$ or $p = 1/2$. The distribution (10) has the following asymptotic behavior:

$$P(V = n) \propto n^{-3/2} \exp(-C_1 n), \quad n >> 1, \qquad (11)$$

where $C_1 = -\ln(4pq) \geq 0$. Such asymptotic with the factor $n^{-3/2}$ is typical for the Galton-Watson models (Harris, 1963) and, consequently, for any ETAS(F) model with $\lambda(m) = const$. The generating function of the distribution (11) is

$$Ez^V = (1 + 2\Lambda - \sqrt{1 - 4(1+\Lambda)\Lambda(z-1)})/(2\Lambda z), \qquad (12)$$

where $EV = \Lambda = \lambda/(1-\lambda) = p/(1-2p)$ (see (4)).

It is obvious, that (12) cannot be represented as (7). Therefore, the analogue of ETAS* model has the geometric distribution for the $\Delta$-productivity $\nu_\Delta$, but does not have any RT property for the total $\Delta$-productivity $V_\Delta$.

## 3 THE TOTAL $\Delta$ – PRODUCTIVITY

We will consider the ETAS(F) model in a subcritical regime: $\Lambda_1 = E\lambda(m) < 1$. $E\nu(m) = \lambda(m)$ is a non-decreasing, finite, positive function; the other components, namely $f_1(m), f_2(t)$ and $f_3(g|m)$, are free from additional restriction.

As above $m_0 > 0$ is the magnitude of an initial cluster event and $V_\Delta(m_0)$ is the total $\Delta$–productivity of $m_0$. Any initial event can be *fixed* or *random*, namely it has the $f_1(m)$ distribution. A further distinction concerns the *dominant* initial events, i.e. those initial events having maximum magnitude within their cluster. The case of a random $m_0$ is important because identifying the initial cluster event is an unstable practical procedure.

**Statement 2.** *Consider the specified ETAS (F) model. Assume that all moments of F-distribution are finite and F has the RT property.*

Productivity within the ETAS seismicity model

*a) Let $V_\Delta(m_0)$ be the size of the cluster with a fixed or random initial magnitude $m_0 > \Delta > 0$, and $N_{V_\Delta}$ be the number of finite moments of $V_\Delta(m_0)$. Then for any $\Delta$, $N_{V_\Delta}$ coincides with the number $N_\lambda$ of finite integrals $I_n = \int_0^\infty \lambda^n(m) f_1(m) dm$, i.e. $N_{V_\Delta} = N_\lambda = \max\{n : I_n < \infty\}$;*

*b) Let F be Poisson or Geometric distribution and $m_0$ is fixed and dominant, Then $N_{V_\Delta} = \infty$ regardless of the $N_\lambda$-value. This is also true for the random dominant $m_0$-event if additionally*

$$\lambda(m) \int_0^\Delta f_1(m-u) du \le C . \tag{13}$$

**Remark 4**. Statement 2 allows us to judge on the tail attenuation of the $V_\Delta$ distribution because, by the Chebyshev-Markov inequality, $P(\xi \ge u) \le E\xi^n / u^n$ for any random variable $\xi \ge 0$ and $u > 0$. Distributions having all moments will be called *distributions* with *light tails*. Otherwise, we will talk about *heavy tails*.

Let's consider clusters with fixed or random initial magnitude. The conditions of the Statement 2 are easily verified for the regular ETAS components: $\lambda(m) = \lambda e^{\alpha m}, \alpha > 0$ and $f_1(m) = \beta \exp(-\beta m), \beta > \alpha$. In this case we have $N_\lambda < \beta/\alpha$. This means that the total $\Delta$ – productivity distribution has only a heavy tail for small $\beta/\alpha$. The data of Zalapin and Ben-Zion (2016) demonstrate the possibilities of this kind. In the considered case Saichev et al. (2005) give more information on the tail behavior of $V_\Delta$ distribution: if $\beta/\alpha$ is large then the tail is sub exponential and looks like (11).

The case $\alpha = 0$ corresponds to the analogue of the ETAS* model; here $N_\lambda = \infty$ in accordance with (10).

Given $\alpha \le \beta$, the condition (13) is satisfied. This guarantees the light tails in the distribution of the total $\Delta$ – productivity of main shocks. All the observations mentioned above support this fact.

**Remark 5.** The difference in the tail behavior of $V_\Delta$ statistics can be partially explained as follows. The total $\Delta$ – productivity of the main shock with a *fixed* magnitude $m_0$ is based on events in the finite range of magnitudes $(0, m_0)$. In this case, the frequency-magnitude law $f_1(m)$ is replaced by its truncated counterpart in the interval $(0, m_0)$, where all moments of $\lambda(m)$ are finite. Since $N_V = N_\lambda$, the tail of the $V_\Delta$ distribution becomes typical for the case $N_\lambda = \infty$ that is quasi exponential. The similar effect for a random dominant magnitude is not so obvious and is a non-trivial analytical fact.

G. Molchan, E. Varini, A. Peresan

**Statement 3.** 1) *Let $m_0$ is a dominant initial magnitude in the ETAS(G) cluster and $f_1(m)$ is strictly positive. Then the statistics $V_\Delta(m_0)$ can have a geometric distribution for no more than one value $\Delta \in (0, m_0)$.*

*2) Let $\nu$ and $V$ be the sizes of clusters of direct descendants and all descendants, respectively, in the ETAS(F) cluster with the random initial event. Assume that both distributions of $\nu$ and $V$ have the RT property, as well as $E\nu^2 < \infty$ and $\int \lambda^2(m) f_1(m) dm < \infty$. Then the distribution types of $\nu$ and $V$ cannot be the same.*

**Remark 6.** According to Statement 3 we can say that, in the framework of generalized ETAS model, the geometric laws for $\nu_\Delta$ and $V_\Delta$, including the case of aftershocks, are incompatible. In the ETAS(G) model case with $\lambda(m) = const$, the geometric law of $\nu_\Delta$ completely excludes the RT property for $V_\Delta$ (see (12)). The reason for this lies in the branching structure of the cluster.

*Numerical examples*

To support the theoretical conclusions, we simulated a synthetic earthquake catalog from the regular ETAS(P) model. To mimic the features of real seismicity, the set of ETAS parameters was mainly derived from the analysis of the earthquake catalog of Northeastern Italy (Benali et al., 2020). The simulation region also corresponds to that area, namely: Lon 11.5-14.0 and Lat 45.5-47.0.

The model was based on the following parameters:
- the uniform background seismicity with the intensity $\mu = 0.04$ (events/($degree^2 \times day$)), which roughly corresponds to the rate of background events identified in the earthquake catalog of Northeastern Italy. The uniformity makes it easier the declustering process, as it allows to better identify the clustered component;
- magnitude range m>2; the Gutenberg-Richter parameter $\beta = 2.07$ which corresponds to the b-value 0.9;
- the Utsu Law parameters: $\lambda = 0.22 (event/day)$, $\alpha = 1.54$. The estimated $\lambda = 0.67$ for Northeastern Italy was reduced to 0.22 to satisfy the stability conditions $\Lambda_1 = E\lambda(m) < 1$;
- the Omori law parameters: $c = 0.015 (day)$, $p = 1.037$. The estimated $p$ parameter very close to 1 implies an extremely long tail of $f_2(t)$, so that offspring events may span several thousands years (Harte, 2013); this practically hinders the analysis of cluster size for synthetic catalogs. However, we recall that, from the theoretical viewpoint, the temporal component of the ETAS model does not affect cluster size. Therefore, in order

Productivity within the ETAS seismicity model

>     to prevent this inconvenience, instead of increasing the $p$ parameter, we considered a truncated distribution $f_2(t)$ on the support $t < 0.7$ years, which corresponds to the 95% quantile of the cluster lifetime for North- eastern Italy;

- the space scattering law parameters: $d = 0.0085\,(degree)$; $q = 2.25$, $\gamma = 0.62$.

The synthetic earthquake catalog, consisting of 143568 events, spans about 500 years and includes information about the links between each event and its direct descendants. The synthetic time series were processed by the Nearest-Neighbor method (Zaliapin & Ben-Zion 2013) to identify clusters of events. This method allows partitioning earthquakes into background and clustered components, based on nearest-neighbor distances between earthquakes in the space–time–magnitude domain (Baiesi & Paczuski 2004; Zaliapin et al. 2008). In this study, the parameters necessary for the computation of nearest neighbor distances, namely the b-value and the fractal dimension of epicenters $d_f$, were set according to the values estimated from the analysis of Northeastern Italy catalog (Benali et al. 2020, and references therein): $b-value = 0.9$ and $d_f = 1.1$. The threshold of the nearest neighbor distance, which separates the clustered and background components, was automatically set ($\log_{10}(\eta_0) = -4.2$) following a criterion based on a one-dimensional Gaussian mixture model with two modes, where the threshold is the maximum likelihood boundary between the two modes (Zaliapin & Ben-Zion, 2013). The total $\Delta$-productivity distribution was examined considering the clusters extracted from the synthetic earthquake catalog.

Fig. 1 represents the total $\Delta$ – productivity distributions ($\Delta = 2$) a) for arbitrary initial events and b) for main shocks, as identified by nearest-neighbor method applied to the synthetic ETAS($P$) seismicity. In the first case, the distribution $V_\Delta$ demonstrates a heavy tail, which is also confirmed by the coefficient of variation $\sigma(V_\Delta)/EV_\Delta = 10.8$ On the contrary, the distribution $V_\Delta$ for main shocks (Fig1b) demonstrates the light tail behavior; for comparison with the previous case $\sigma(V_\Delta)/EV_\Delta = 0.7$.

Fig.2 shows similar results based on the clusters originally defined by the branching structure of the synthetic ETAS($P$) catalog. The coefficients of variation are also consistent with previous results: (a) $\sigma(V_\Delta)/EV_\Delta = 9.7$ for arbitrary initial events and (b) $\sigma(V_\Delta)/EV_\Delta \approx 0.8$ for main shocks.

A good agreement of the empirical data in Fig. 1 and Fig. 2 demonstrates the robustness of conclusions about the tail behavior of $V_\Delta$ distributions.

G. Molchan, E. Varini, A. Peresan

**(a)** The empirical data: $N = 612 \quad EV_\Delta = 39.8 \quad \sigma(V_\Delta) = 431.3 \quad \max V_\Delta = 10079$

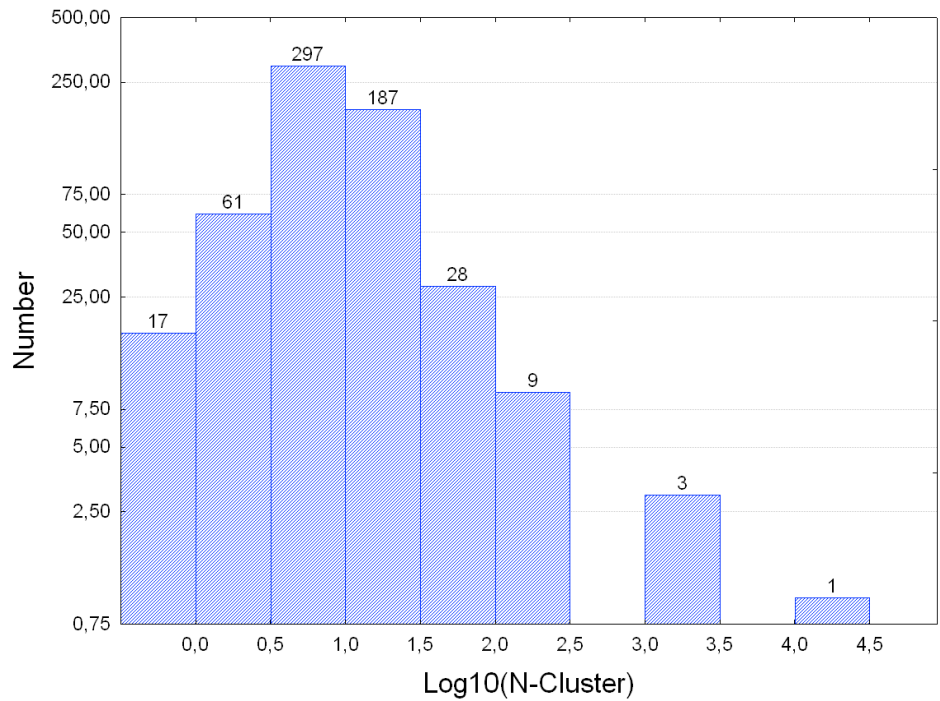

**(b)** The empirical data: $N = 1074 \quad EV_\Delta = 9.2 \quad \sigma(V_\Delta) = 6.5 \quad \max V_\Delta = 52$

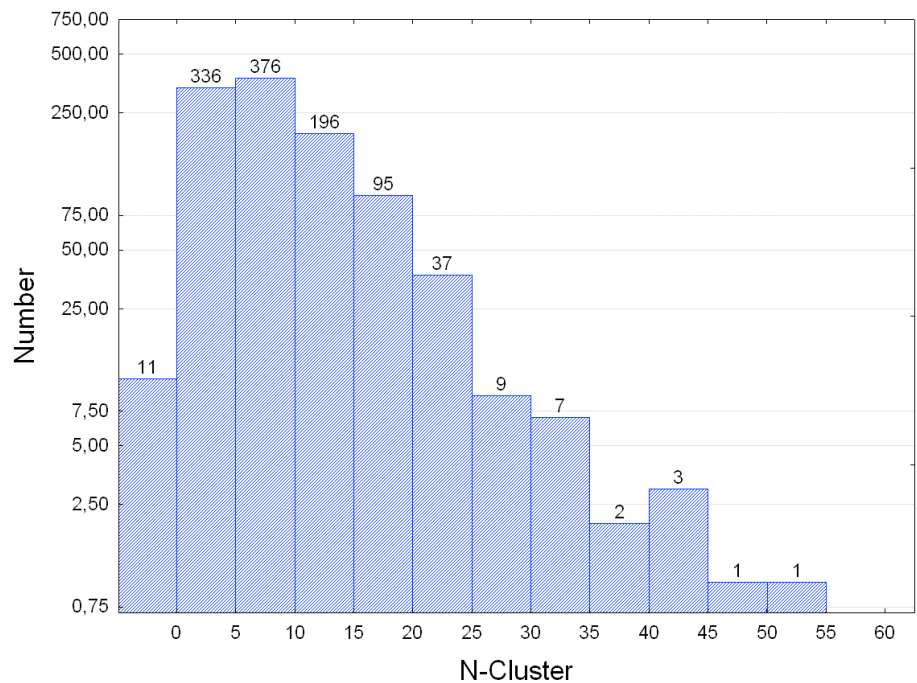

**Figure 1.** Clusters identified by nearest-neighbor method applied to the synthetic (ETAS($P$)) seismicity: $V_\Delta, \Delta = 2$ distributions for a) arbitrary initial events, $m_0 \geq 4$ and b) main shocks, $m_0 \geq 4$.

Productivity within the ETAS seismicity model

**(a)** The empirical data: $N = 481$  $EV_\Delta = 46.5$  $\sigma(V_\Delta) = 452.3$  $\max V_\Delta = 9470$

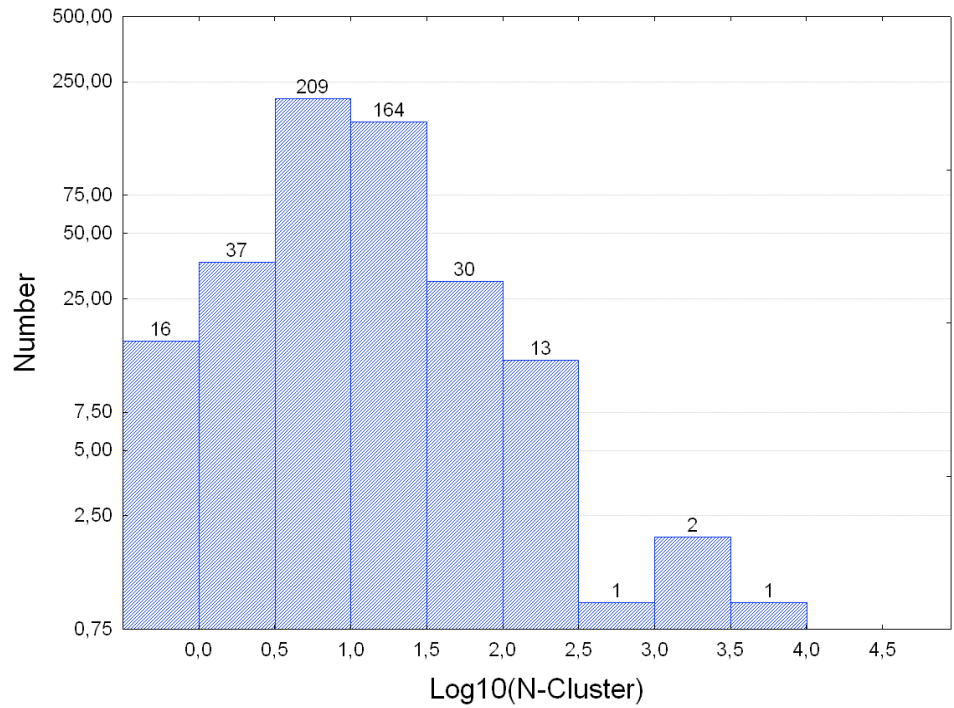

**(b)** The empirical data: $N = 1074$  $EV_\Delta = 9.9$  $\sigma(V_\Delta) = 7.9$  $\max V_\Delta = 56$

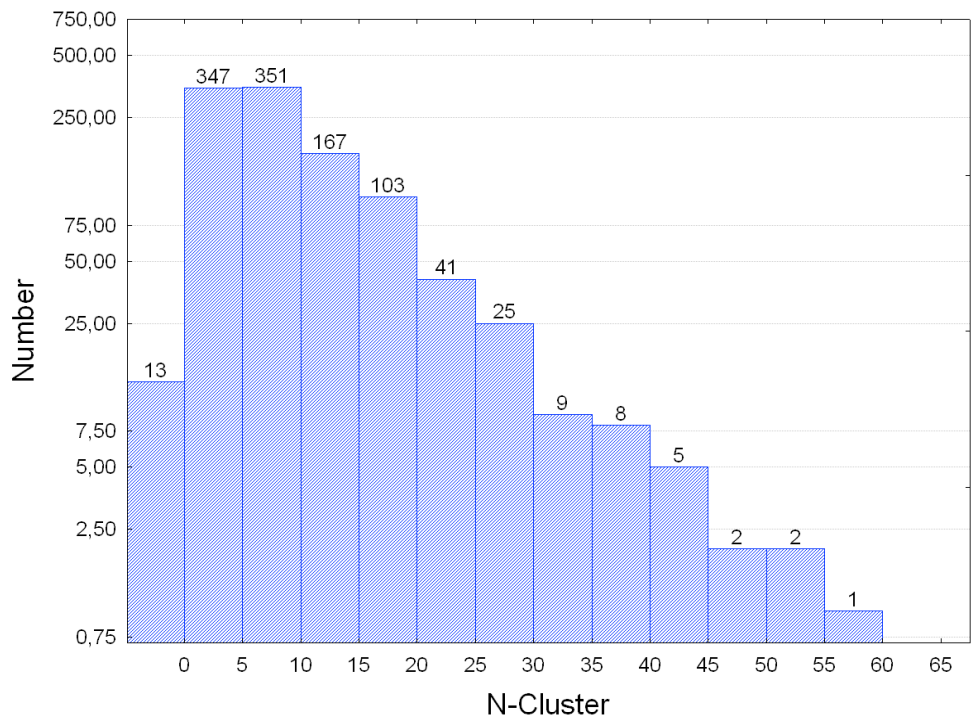

**Figure 2**. Clusters in the synthetic (ETAS(*P*)) seismicity: $V_\Delta, \Delta = 2$ distributions for a) arbitrary initial events, $m_0 \geq 4$ and b) main shocks, $m_0 \geq 4$.

G. Molchan, E. Varini, A. Peresan

## 4 CONCLUSION

The concept of the $\Delta$-productivity $\nu_\Delta$ is associated with a priori ideas about the structure of a cluster of seismic events in the form of a random branching tree. Such a representation is not the only possible one. Therefore, the Geometric distribution of $\nu_\Delta$ can be considered as a useful alternative within the framework of the generalized ETAS(F) model. The other characteristic of the $\Delta$-productivity $V_\Delta$ for main shocks is more objective and stable.

For large class of the ETAS(F) models, we answered the question when the total $\Delta$ productivity distribution has light tails for main shocks and heavy ones for arbitrary events. The real data demonstrate the possibilities of this kind.

Note the difficulties of substantiating the mentioned geometric laws for $\nu_\Delta$ and aftershock's $V_\Delta$. They are associated both with the declustering of seismicity and with the effect of averaging the distributions of $\nu_\Delta$ and $V_\Delta$ when grouping data.

The first difficulty can be judged by experiments of Zaliapin and Ben-Zion (2013) with the Nearest-Neighbor method. This method adapted to the ETAS structure has 40% error of incorrectly determining the parent of an event. The stability of the conclusions in this case may depend on the strict similarity in the real hierarchical structure of the seismic cluster.

The second difficulty is expressed in the fact that the mixture of geometric distributions retains the original property of monotonicity, but ceases to be geometric, i.e. loses the original G-type. This is unavoidable if the distribution parameter is dependent on location or the magnitude of the initial event (see (6)). Therefore, we can speak more confidently about the light tail of the $V_\Delta$- distribution for the main shocks.

Finally, the theoretical result for ETAS(F) model shows that the types of distributions of $\nu_\Delta$ and $V_\Delta$ cannot be the same due to the branching structure of the clusters.

**REFERENCES**

Baiesi, M., & Paczuski, M., 2004. Scale-free networks of earthquakes and aftershocks. *Physical Review E*, **69,** 066106.

Baranov S., Narteau C., Shebalin P., 2022. Modeling and Prediction of Aftershock Activity. *Surveys in Geophysics* .DOI:10.1007/s10712-022-09698-0

Baranov S.V., Shebalin P.N., 2019. *Laws of the post-seismic process and the forecast of strong aftershocks* .Moscow, RAS, 218 pp.

Baro J., 2020.Topological Properties of Epidemic Aftershock Processes. *J.Geophys.Res.:Solid Earth* ,**125**:5 .

Productivity within the ETAS seismicity model

Benali, A., Peresan, A., Varini, E., & Talbi, A. ,2020. Modelling background seismicity components identified by nearest neighbour and stochastic declustering approaches: The case of northeastern Italy. *Stoch. Environ. Res. Risk. Assess*. **34**, 775–791.

Harris T.E., 1963.*The theory of branching processes*. Springer-Verlag, Berlin.

Harte, D. S., 2013. Bias in fitting the ETAS model: a case study based on New Zealand seismicity, *Geophys. J. Int*., **192**(1), 390–412.

Hawkes A.G., Oakes D .,1974. A cluster process representation of a self-exciting process. *Journal of Applied Probability* **11**(3):493-503.

Kagan Y., 2010. Statistical distributions of EQ numbers: consequence of brunching process, *Geophys. J. Int.,* **180**(3), 1313-1328.

Kagan Y.,2014. *Earthquakes: Models,Statistics, Testable Forecasts*, pp.283,AGU, Wiley .

Ogata, Y., 1998. Space- time point- process models for earthquake occurrences. *Annals of the Institute of Statistical Mathematics*, **50**(2), 379–402.

Saichev, A., Helmstetter, A., & Sornette,D., 2005. Power-law distributions of offspring and generation numbers in branching models of earthquake triggering. *Pure and applied geophysics,* **162**(6), 1113–1134.

Saichev A. and Sornette D., 2017. Statistics of seismic cluster durations. *arXiv:1708.06970* .

Shebalin P.N., Baranov S.V., Dzeboev B. A. ,2018. The Law of the Repeatability of the Number of Aftershocks. *Doklady Akademii Nauk,* **481**(3). DOI: 10.1134/S1028334X18070280

Shebalin P.N., Narteau C. and Baranov S.V.,2020. Earthquake productivity law. *Geophys. J. Int.* **222,** 1264–1269.

Shebalin P.N., Baranov S.V.,2021. Statistical laws of post-seismic activity. In book: *Statistical Methods and Modeling of Seismogenesis ,*(Editors: Limnios N.**,** Papadimitriou**,E.,** Tsaklidis G.), 330 pp.

Soloviev S.L and Solovieva O.N., 1962. Exponential distribution of the total number of subsequent earthquakes and decreasing with the depth of its average value. *Izv. AN USSR (geophys.)* **12**, 1685-1694.

Zaliapin, I., Gabrielov, A., Keilis Borok, V., Wong, H. 2008. Clustering analysis of seismicity and aftershock identification. *Phys.Rev.Lett*.,**101**,018501.

Zaliapin I. Ben-Zion Y. , 2013.Earthquake clusters in southern California I: identification and stability. *J. Geophys. Res*. **118,** 2847–2864.

Zaliapin I., Ben-Zion Y.,2016. A global classification and characterization of earthquake clusters. *Geophys. J. Int,* **207**(1), 608–634**.**

G. Molchan, E. Varini, A. Peresan

**APPENDIX: proofs**

**Statement 1**

Let $E(z^{\nu(m_0)}|m_0) = \varphi(z)$ be the generating function of $\nu(m_0)$ with the characteristic $\lambda(m) = E\nu(m)$. If $\{m_i, i = 1,...,\nu(m_0)\}$ are magnitudes of the direct descendants of the $m_0$ event, then $\nu_\Delta(m_0)$ has the same distribution as $\sum_1^{\nu(m_0)} \varepsilon(m_i)$, where $\varepsilon(m) = [m > m_0 - \Delta]$ is the logical 1-0 function. Since $\{m_i\}$ are independent

$$\varphi_\Delta(z) = E(z^{\nu_\Delta(m_0)}|m_0) = \sum_{n \geq 0}(Ez^{\varepsilon(m)})^n P(\nu(m_0) = n) = \varphi(\pi_{m_0-\Delta} + \bar{\pi}_{m_0-\Delta}z) \quad (A.1)$$

where $\bar{\pi}_{m-\Delta} = P(m \geq m_0 - \Delta | m_0 \geq \Delta)$ and $\pi_m = 1 - \bar{\pi}_m$.

Recall that $\varphi(z) = (1-p)/(1-pz) = [1 - \lambda(z-1)]^{-1}$ for F=G and $\varphi(z) = \exp(\lambda(z-1))$ for F=P. Substituting any of the functions $\varphi(z)$ in the right part of (A.1), we get the same function with the replacement of $\lambda(m_0)$ by

$$\lambda_\Delta(m_0) = \lambda(m_0)\bar{F}_1(m_0 - \Delta)/\bar{F}_1(\Delta). \quad (A.2)$$

Assume that the right part of (A.2) is independent of $m_0$ in the interval $\Delta \leq m_0 < M$. Then $\lambda(m)$ and $F_1(m)$ are differentiable functions simultaneously. By (A.2)

$$0 = (d/dm_0)\lambda_\Delta(m_0)|_{m_0=\Delta} = [\dot{\lambda}(\Delta) - \lambda(\Delta)f_1(0)]/\bar{F}_1(\Delta).$$

Here we used condition $\bar{F}_1(0) = 1$. As a result, we get $\lambda(m) = ce^{am}$ and $\bar{F}_1(m) = e^{-am}$, $m \in (0, M)$, where $a = f_1(0)$.

**Statement 2**

**General remarks**. We will need the following notation: $N_M(m)$ is the total number of events of magnitude $\geq M$ triggered by an event of magnitude $m$. In particular, the total $\Delta$-productivity $V_\Delta(m_0)$ of the initial event with magnitude $m_0$ is $N_{M(m_0)}(m_0)$, $M(m) = (m - \Delta)_+$, where $a_+ = \max(a, 0)$. The notation $E(*|m)$ denotes the conditional mean given $m$.

The spatial-temporal components of the ETAS model do not affect cluster size, since the law of generating new events depends only on the parent magnitude, and the problem of de-clustering does not arise in our theoretical analysis. Therefore, the spatial-temporal components in theoretical analysis can be any and the distribution of $N_M(m)$ is the same for any $x = (t, g, m)$.

Productivity within the ETAS seismicity model

By definition, if $\{m_i, i=1,...,\nu(m_0)\}$ are offspring magnitudes of $m_0$ then $\{N_M(m_i), i=1,...,\nu(m_0)\}$ are independent for a given $m_0$. Moreover, they are identically distributed together with $N_M(m_0)$ since the magnitudes $m_i, i \geq 0$ have a common distribution $f_1(m)$. Let us consider a new variable

$$\tilde{N}_M(m_0) = \sum_{1}^{\nu(m_0)} \{N_M(m_i) + [m_i \geq M]\}, \tag{A.3}$$

where the symbol $[a > b]$ denotes the logical function 1-0. Obviously, (A.3) is nothing more than counting events with magnitude $\geq M$ in the cluster through first-generation events and its triggered events of magnitude $\geq M$. The component $[m_i \geq M]$ takes into account i-th event of the first generation if its value is greater than $M$. Therefore $N_M(m_0)$ and $\tilde{N}_M(m_0)$ are identically distributed, that is $N_M(m_0) =_{law} \tilde{N}_M(m_0)$.

**The moments** $EV_\Delta^n$.

We will use the notation:

$$\Lambda_n = \int \lambda^n(m) f_1(m) dm, \quad \overline{\Lambda}_1 = 1 - \Lambda_1, \quad \lambda(m) = p(m)/(1-p(m)) \text{ in the case F=G}.$$

$$b_M^{(n)}(m) = E[N_M^n(m)|m], \quad b_M^{(n)} = EN_M^n(m), \quad \pi_M = E[m > M].$$

*Mean value:* $EV_\Delta$. Averaging (A.3) given $m_0 = m$ we have

$$b_M^{(1)}(m) = \lambda(m)(b_M^{(1)} + \pi_M).$$

After averaging over $m$, we obtain the equation for $b_M^{(1)}$:

$$b_M^{(1)} = \Lambda_1(b_M^{(1)} + \pi_M).$$

For finite $b_M^{(1)}$, this equality can be true, if $\Lambda_1 < 1$. Hence

$$b_M^{(1)} = \pi_M \cdot \Lambda_1 / \overline{\Lambda}_1, \quad b_M^{(1)}(m) = \lambda(m)\pi_M / \overline{\Lambda}_1. \tag{A.4}$$

Since $V_\Delta = N_{M(m)}(m)$, $M(m) = (m-\Delta)_+$, we have

$$E(V_\Delta|m) = \lambda(m)\overline{F}_1(m-\Delta)/\overline{\Lambda}_1, \quad E(V_\Delta) \leq \Lambda_1/\overline{\Lambda}_1,$$

$$E(V_\Delta|m \geq \Delta) \leq (\Lambda_1/\overline{\Lambda}_1)/\overline{F}_1(\Delta),$$

that is, $V_\Delta$ has finite mean value in conditional and unconditional situations if $\Lambda_1 < 1$.

*The case* $n > 1$.

**a)** *Fixed/random initial* $m_0$

It is easy to see that for any $\lambda(m) \geq 0$ there exists $K \leq \infty$ such that $\Lambda_k = E\lambda^k(m)$ is finite for $k \leq K$ and is infinite for $k > K$. Let us show that

$$b_M^{(k)} \leq B_k, \quad b_M^{(k)}(m) < \infty \text{ if and only if } \Lambda_k < \infty, \Lambda_1 < 1 \tag{A.5}$$

G. Molchan, E. Varini, A. Peresan

We have shown (see above) that (A.5) is true for $k = 1$. We will prove (A.3) by the method of induction on $k$. Assume that all values in (A.5) are finite for $k \leq n - 1$.

Putting $\xi_i = N_M(m_i) + [m_i \geq M]$ in (A.3), one has

$$N_M^n(m) = \sum\nolimits_{i_1 \neq i_2 \neq \ldots \neq i_n}^{\nu(m)} \xi_{i_1} \ldots \xi_{i_n} + R_M^{(n)} + \sum\nolimits_{i=1}^{\nu(m)} \xi_i^n. \tag{A.6}$$

The number of terms in the first sum is $\eta_n = \nu(m)(\nu(m) - 1)\ldots(\nu(m) - n + 1)$.

By the RT property of the F- distribution, $Ez^{\nu(m)} = \varphi(z,m) = \psi(\lambda(m)(z-1))$, where $\varphi(z,m)$ has all derivative at z=1 because all moments of $\nu(m)$ are finite. Hence

$$E[\eta_n | m] = (d/dz)^n \varphi(z,m)\big|_{z=1} = a_n(F)\lambda^n(m), \tag{A.7}$$

where $a_n(F) = (d/dz)^n \psi(z)\big|_{z=0} < \infty$.

Therefore

$$E(\sum\nolimits_{i_1 \neq i_2 \neq \ldots \neq i_n}^{\nu(m)} \xi_{i_1} \ldots \xi_{i_n} | m) = a_n(F)\lambda^n(m)[b_M^{(1)} + \pi_M]^n.$$

Similarly

$$E(R_M^{(n)} | m) = \sum\nolimits_\Omega c_n(\omega) a_k(F)\lambda^k(m)\tilde{b}_M^{(r_1)} \ldots \tilde{b}_M^{(r_k)}, \tag{A.8}$$

$$\tilde{b}_M^{(r)} = E\{N_M(m_i) + [m_i \geq M]\}^r,$$

where

$$c_n(\omega) = n!/(k_1!\ldots k_r!r!), \quad \Omega = \{\omega = (k, r_1, \ldots, r_k) : 1 < k < n, r_i \geq 1, r_1 + \ldots + r_k = n\}.$$

Since for any random variable $\xi \geq 0$:

$$E\xi^n E\xi^m \leq E\xi^{n+m},$$

we get from (A.8):

$$E(R_M^{(n)} | m) \leq C_n (1 \vee \lambda(m))^{n-1} \max\nolimits_{1 \leq k < n} (\tilde{b}_M^{(k)} \tilde{b}_M^{(n-k)}). \tag{A.9}$$

The expectation of (A.6), given $m$, is

$$b_M^{(n)}(m) = a_n(F)\lambda^n(m)(\tilde{b}_M^{(1)})^n + E(R_M^{(n)} | m) + \lambda(m)[b_M^{(n)} + \delta_M^{(n)}], \tag{A.10}$$

where $b_M^{(n)} + \delta_M^{(n)} = \tilde{b}_M^{(n)}$. By (A.8),

$$\delta_M^{(n)} < E(N_M(m) + 1)^n - N_M^n(m)]$$
$$< nE(N_M(m) + 1)^{n-1} < n(2^{n-1} b_M^{(n-1)} + 1). \tag{A.11}$$

Averaging (A.10) over $m$, we get

$$b_M^{(n)}(1 - \Lambda_1) = a_n(F)\Lambda_n(\pi_M \Lambda_1 / \bar{\Lambda}_1)^n + ER_M^{(n)} + \Lambda_1 \delta_M^{(n)}. \tag{A.12}$$

Productivity within the ETAS seismicity model

The relations (A.11, A.12) show that $b_M^{(n)}$ is finite if and only if $\Lambda_n < \infty$. Similarly, (A.10, A.11) show that $b_M^{(n)}(m) < \infty$ if ad only if $b_M^{(n)} < \infty$. Hence, at the n-th step of induction, these three quantities are finite simultaneously, that is (A.5) holds.

By setting $M = (m-\Delta)_+$, we get: $E(V_\Delta^n | m) = b_{(m-\Delta)_+}^{(n)}(m) < \infty$ only when $\Lambda_n < \infty$.

It remains to consider $E(V_\Delta^n)$. By (A.10, A.12),

$$b_M^{(n)}(1-\Lambda_1) = a_n(F)\Lambda_n(\pi_M \Lambda_1 / \bar{\Lambda}_1)^n + ER_M^{(n)} + \Lambda_1 \delta_M^{(n)}$$

$$E(V_\Delta^n) = Eb_{(m-\Delta)_+}^{(n)}(m) \geq E\lambda(m)b_{(m-\Delta)_+}^{(n)} \geq E\lambda(m)a_n(F)\Lambda_n \pi_{(m-\Delta)_+}^n \Lambda_1^n / \bar{\Lambda}_1^{n+1}.$$

Therefore $E(V_\Delta^n) = \infty$ if $\Lambda_n$ is infinite.

Let $\Lambda_n < \infty$. Using (A.10) again, we have

$$EV_\Delta^n \leq a_n(F)\Lambda_n(\Lambda_1/\bar{\Lambda}_1)^n + ER_{M(m)}^{(n)}(m) + E\lambda(m)[b_{M(m)}^{(n)} + \delta_{M(m)}^{(n)}].$$

Due to (A.9), we have $ER_{M(m)}^{(n)}(m) \leq \widetilde{C}_n \max(1, \Lambda_1, ... \Lambda_{n-1})$; by (A.9), $E\lambda(m)\delta_{M(m)}^{(n)} \leq D_n \Lambda_1$. Finally, $E\lambda(m)b_{M(m)}^{(n)} \leq B_n \Lambda_1$, because we have proved (A.3) for index $n$. Hence, $E(V_\Delta^n) < \infty$ if $\Lambda_n < \infty$. The proof is complete.

**b) Fixed dominant $m_0$**

Let us consider the ETAS cluster with a dominant initial magnitude $m_0$. Let $\hat{N}_{0M}(m)$ be the number of events of magnitude $> M$ in the cluster with the initial event $m \leq m_0$ in which all events have magnitude $< m_0$. In such cluster, direct descendants of any initial magnitude $m$ are realized under the condition $\Omega(m) = \{m_i < m_0, i = 1,..,\nu(m_0)\}$.

In term of the generating function $Ez^{\nu(m)} = \varphi(z,m)$, the probability of $\Omega(m)$ is

$$P(\Omega(m)) = \varphi(F_1(m_0), m) = \psi(-\lambda(m)\bar{F}_1(m_0))  \qquad (A.13)$$

Therefore the conditional probability density of the direct descendants is

$$P\{m_i = \mu_i, i = 1,..,n, \nu(m_0) = n | \Omega(m)\} = \hat{f}_{01}(\mu_1)...\hat{f}_{01}(\mu_n)P(\hat{\nu}(m) = n).$$

where

$$\hat{f}_{01}(m) = f_1(m)/F_1(m_0), 0 \leq m \leq m_0 \quad F_1(m) = \int_0^m f_1(u)du \qquad (A.14)$$

$$P(\hat{\nu}(m) = n) = P(\nu(m) = n | \Omega(m)) = P(\nu(m) = n)F_1^n(m_0)/\psi(-\lambda(m)\bar{F}_1(m_0)) \quad (A.15)$$

Substituting Geometric (1) and Poisson (2) distributions here, we get the distributions of the same type

$$P\{\nu(m_0) = n | \Omega(m)\}$$

G. Molchan, E. Varini, A. Peresan

$$= \begin{cases} \hat{\lambda}_0^n(m) e^{-\hat{\lambda}_0(m)} / n! & P-\text{model} \\ \hat{p}_0^n(m)(1-\hat{p}_0(m)) & G-\text{model} \end{cases}, \quad (A.16)$$

where

$$\hat{p}_0(m) = p(m) F_1(m_0) = \lambda(m) F_1(m_0)/(1+\lambda(m)) \ (F=G),$$

$$\hat{\lambda}_0(m) = \lambda(m) F_1(m_0) \begin{cases} 1 & , F = P \\ (1+\lambda(m) \overline{F}_1(m_0))^{-1} & , F = G \end{cases} \quad (A.17)$$

Therefore, given the dominance of the initial magnitude $m_0$, we are again dealing with an ETAS model in which the magnitude $m$ is transformed into $\hat{m}$ with a distribution (A.14), and the productivity $\nu(m)$ into $\hat{\nu}(\hat{m})$ with the same distribution (Poisson/Geometric), but with new parameters $\hat{\lambda}_0(\hat{m})$ as in (A.17). Therefore, as above, we will have the following relations

$$\hat{N}_{0M}(\hat{m}) =_{law} \sum_{i=1}^{\hat{\nu}(\hat{m})} (\hat{N}_{0M}(\hat{m}_i) + [\hat{m}_i - M]), \quad (A.18)$$

where $\{\hat{N}_{0M}(\hat{m}_i), i = 1,...\}$ are independent and identical distributed given $\{\hat{m}_i, i \geq 0\}$.

As a result

$$\nu_\Delta(m_0) = \hat{N}_{0(m_0-\Delta)}(m_0), \quad m_0 \geq \Delta. \quad (A.19)$$

Since $m_0$ is a parameter, in the random case of initial event, $m_0$ has distribution $F_1(m)/\overline{F}_1(\Delta)$.

As we can see, equation (A.18) is a special case of (A.3), in which the magnitude range is finite. By (A.17), $\hat{\lambda}_0(\hat{m}) \leq \lambda(m_0)$ and therefore $\hat{\lambda}_0(\hat{m})$ has all moments. Consequently, the total $\Delta$-productivity will have all the moments at any fixed $m_0$.

**c) Random dominant** $m_0$. We need additional notation:

$$\hat{b}_{0M}^{(n)}(\hat{m}) = E[\hat{N}_{0M}^n(\hat{m})|\hat{m}], \quad \hat{b}_{0M}^{(n)} = E\hat{N}_{0M}^n(\hat{m}),$$

$$\tilde{b}_{0M}^{(n)} = E(\hat{N}_{0M}(\hat{m}) + [\hat{m} > M])^n,$$

$$\hat{\pi}_0(M) = E[\hat{m} > M] = [F_1(m_0) - F_1(M)]_+ / F_1(m_0), \quad (A.20)$$

$$\hat{\Lambda}_{0n} = E\hat{\lambda}_0^n(\hat{m}).$$

Using (A.17) and non-decreasing function $\lambda(m)$, we have for the P- model

$$\hat{\Lambda}_{0n} = \int_0^{m_0} \lambda^n(x) F_1^{n-1}(m_0) f_1(x) dx \leq \hat{\lambda}_0^{n-1}(m_0) \Lambda_1. \quad (A.21)$$

The same is true for the G-model. Really, by (A.17) we have the following representation:

$$\hat{\Lambda}_{0n} = \hat{\lambda}_0^{n-1}(m_0) \int_0^{m_0} [\psi^{n-1}(\lambda)/(1+\lambda(m)\overline{F}(m_0))]\lambda(x) f_1(x) dx,$$

Productivity within the ETAS seismicity model

where
$$\psi(m) = (\lambda^{-1}(m_0) + \overline{F}_1(m_0))/(\lambda^{-1}(m) + \overline{F}_1(m_0)).$$

Since $\lambda(m)$ is a non-decreasing function, $\psi(m) \leq 1$ on the interval $(0, m_0)$. Therefore, the integrand in square brackets is less than 1, which proves (A.21).

As above, the relation (A18) gives,
$$\hat{b}_{0M}^{(1)} = \hat{\Lambda}_{01}\hat{\pi}_0(M)/(1-\hat{\Lambda}_{01}) \leq \Lambda_1 \hat{\pi}_0(M)/(1-\Lambda_1), \tag{A.22}$$
$$E\nu_\Delta \leq \Lambda_1/(1-\Lambda_1).$$

Setting $\xi_i = \hat{N}_{0M}(\hat{m}_i) + [\hat{m}_i - M]$ in (A.18) and using (A.6), we get an analogue of (A.10):
$$\hat{b}_{0M}^{(n)}(\hat{m}) = \sum_{k=1}^{n} \sum_{\mathbf{r}^{(k)}} c_k(r_1,\ldots,r_k) \hat{\lambda}_0^k(\hat{m}) \widetilde{b}_{0M}^{(r_1)} \ldots \widetilde{b}_{0M}^{(r_k)}, \tag{A.23}$$

where the summation is over all integer vectors $\mathbf{r}^{(k)} = (r_1,\ldots,r_k): r_1 + \ldots + r_k = n; r_i \geq 1$. Averaging (A.23) over $\hat{m}$ we will also have
$$\hat{b}_{0M}^{(n)} = \sum_{k=1}^{n-1} \sum_{\mathbf{r}^{(k)}} c_k(r_1,\ldots,r_k) \hat{\Lambda}_{0k} \widetilde{b}_{0M}^{(r_1)} \ldots \widetilde{b}_{0M}^{(r_k)} + \hat{\Lambda}_{01} \widetilde{b}_{0M}^{(n)}. \tag{A.24}$$

Let's extract the term $\hat{\Lambda}_{01}\hat{b}_{0M}^{(n)}$ from $\hat{\Lambda}_{01}\widetilde{b}_{0M}^{(n)}$ and move it to the left part of (A.24); then apply (A.21) to $\hat{\Lambda}_{0k}$ and take into account the relation $\hat{\Lambda}_{01}/(1-\hat{\Lambda}_{01}) \leq \Lambda_1/(1-\Lambda_1)$. Then multiplying (A.24) by $\hat{\lambda}_0(m_0)$, we get
$$\hat{\lambda}_0(m_0)\hat{b}_{0M}^{(n)} \leq [\sum_{0\leq k<n} c_k(r_1,\ldots,r_k)(\hat{\lambda}_0(m_0)\widetilde{b}_{0M}^{(r_1)})\ldots(\hat{\lambda}_0(m_0)\widetilde{b}_{0M}^{(r_k)})$$
$$+ \hat{\lambda}_0(m_0)(\widetilde{b}_{0M}^{(n)} - \hat{b}_{0M}^{(n)})]\Lambda_1/(1-\Lambda_1). \tag{A.25}$$

Since $\widetilde{b}_{0M}^{(n)} = E(\hat{N}_{0M}(\hat{m}_i) + [\hat{m}_i > M])^n$, the elements of the right part of (A.25) can be obtained by summing and multiplying elements of the form: $\hat{\lambda}_0(m_0)\hat{\pi}_0(M)$ and
$$D_{k,\varepsilon} = \hat{\lambda}_0(m_0) E\hat{N}_{0M}^k(\hat{m})[\hat{m} \geq M]^\varepsilon, k=1,\ldots,n-1; \varepsilon = 0,1, \tag{A.26}$$

where $D_{k,\varepsilon} \leq \hat{\lambda}_0(m_0)\hat{b}_{0M}^{(k)}$.

We have assumed that $\lambda(m)\int_0^\Delta f_1(m-x)dx \leq C$. Hence
$$\hat{\lambda}_0(m_0)\hat{\pi}_0(m_0 - \Delta) = \lambda(m_0)F_1(m_0)[F_1(m_0) - F_1(m_0 - \Delta)]/F_1(m_0) \leq C$$

and by virtue of (A.22)
$$\hat{\lambda}_0 \hat{b}_{0M(m_0)}^{(1)} \leq C_1, \quad M(m_0) = (m_0 - \Delta)_+.$$

Using (A.25) and doing induction over n, we can conclude that $\hat{\lambda}_0 \hat{b}_{0M(m_0)}^{(n)} \leq C_n$ for any $n$. But then, due to (A.19), $E\nu_\Delta^n = E\hat{b}_{0M(m_0)}^{(n)}(m_0) \leq K_n < \infty$ for any n.

G. Molchan, E. Varini, A. Peresan

The proof is complete.

## Statement 3.1

Let $\xi$ be a random variable with a geometric distribution. Then

$$E\xi(\xi-1)/(E\xi)^2 = 2. \quad (A.27)$$

Consider $\xi = V_\Delta(m_0)$ where $m_0$ is the initial dominant magnitude in the ETAS(G) cluster. We will show that the relation (A.27) is possible for at most one parameter $\Delta \in (0, m_0)$.

For simplicity of notation, we will temporarily proceed from equation (A.3). Recall that

$$\Lambda_n = \int \lambda^n(m) f_1(m) dm, \quad \pi_M = E[m > M],$$

$$b_M^{(n)}(m) = E[N_M^n(m)|m], \quad b_M^{(n)} = E N_M^n(m).$$

Above we show that

$$b_M^{(1)}(m) = \lambda(m)\pi_M/\overline{\Lambda}_1, \quad b_M^{(1)} = \pi_M \cdot \Lambda_1/\overline{\Lambda}_1, \quad b_M^{(1)} + \pi_M = \pi_M/\overline{\Lambda}_1. \quad (A.28)$$

According to (A.10) for the ETAS(G) model, we have

$$b_M^{(2)}(m) = 2(\lambda(m)\pi_M/\overline{\Lambda}_1)^2 + \lambda(m)(b_M^{(2)} + 2b_M^{(1)}\pi_M + \pi_M^2). \quad (A.29)$$

By (A.28),

$$2b_M^{(1)}\pi_M + \pi_M^2 = (1+\Lambda_1)\pi_M^2/\overline{\Lambda}_1. \quad (A.30)$$

Averaging (A.29) over $m$, we can find $b_M^{(2)}$; using (A.30), we will have

$$b_M^{(2)} = 2\Lambda_2\pi_M^2/\overline{\Lambda}_1^3 + \Lambda_1(1+\Lambda_1)\pi_M^2/\overline{\Lambda}_1^2. \quad (A.31)$$

If we substitute (A.31) into (A.29), we get

$$b_M^{(2)}(m) = 2(b_M^{(1)})^2 + \lambda(m)(\pi_M/\overline{\Lambda}_1)^2(2\Lambda_2 + 1 - \Lambda_1^2)/\overline{\Lambda}_1.$$

As a result,

$$(b_M^{(2)}(m) - b_M^{(1)})/[b_M^{(1)}]^2 = 2 + \lambda^{-1}(m)[(2\Lambda_2 + 1 - \Lambda_1^2)/\overline{\Lambda}_1 - \overline{\Lambda}_1/\pi_M]. \quad (A.32)$$

The condition (A.27) is satisfied only if

$$\pi_M = (1-\Lambda_1)^2/(2\Lambda_2 + 1 - \Lambda_1^2), \quad (A.33)$$

where $\Lambda_2 - \Lambda_1^2 = \sigma^2(\lambda) > 0$.

The equation (A.33) with respect to $M$ has unique solution if $\pi_M$ is strictly monotone.

Now we can return to the ETAS(G) cluster with dominant initial magnitude $m_0$. In this case the main equation is (A.18). It is identical to (A.10), but requires the following replacement of the characteristics used (see (A.14, A.17)):

$$f_1(m) \Rightarrow f_1(m)/F_1(m_0), 0 \le m \le m_0,$$

Productivity within the ETAS seismicity model

$$\lambda(m) \Rightarrow \lambda(m)F_1(m_0)/(1+\lambda(m)\overline{F_1}(m_0)),$$

$$\pi_M \Rightarrow [F_1(m_0) - F_1(M)]_+ / F_1(m_0), \tag{A.34}$$

$$\Lambda_n \Rightarrow \int_0^{m_0} \lambda^n(m) F_1^{n-1}(m_0)(1+\lambda(m)\overline{F_1}(m_0))^{-n} f_1(m)dm. \tag{A.35}$$

Since $V_\Delta(m_0) = N_{m_0-\Delta}(m_0)$, we have to consider equation (A.33) given the substitutions (A34, A35) and $M = m_0 - \Delta$. Given $f_1(m) > 0$ the function (A.34) will be strictly monotonic with respect to $\Delta$. Hence, the relation (A.27) is possible for at most one value of $\Delta = \Delta(m_0)$.

The proof is complete.

## Statement 3.2

Let $v(m)$ and $V(m)$ be the sizes of clusters of direct descendants and all descendants, respectively, associated to an initial event with magnitude $m$. Let $\varphi(z|m) = Ez^{V(m)}$. Due to the RT property of $v(m)$ distribution, we have

$$Ez^{v(m)} = \psi(\lambda(m)(z-1)), \psi(0) = \psi'(0) = 1, \psi''(0) < \infty.$$

Let $\varphi(z) = Ez^V = \int_0^\infty \varphi(z|m) f_1(m)dm$. Then by (A.3)

$$\varphi(z|m) = \psi(\lambda(m)(z\varphi(z) - 1)).$$

and after integration over $m$ we get

$$\varphi(z) = \int_0^\infty \psi(\lambda(m)(z\varphi(z) - 1)) f_1(m)dm. \tag{A.36}$$

As above, $EV^2 < \infty$ if $\int \lambda^2(m) f_1(m)dm < \infty$ and $Ev^2 < \infty$. This means that $\varphi''(1) < \infty$.

Set $\mu = EV = \varphi'(1)$, $\Lambda_k = \int_0^\infty \lambda^k(m) f_1(m)dm$. Differentiating (A.36) at point 1, we get

$$\mu = \Lambda_1(1+\mu) \text{ or } \mu = \Lambda_1/(1-\Lambda_1). \tag{A.37}$$

Differentiating (A.36) twice at point 1, we get

$$\varphi''(1) = \psi''(0)\Lambda_2(1+\mu)^2 + \Lambda_1(2\mu + \varphi''(1)).$$

Given (A.37), we have

$$\varphi''(1) = \psi''(0)\Lambda_2(1-\Lambda_1)^{-3} + 2\mu^2 \tag{A.38}$$

Assume that the distribution types of $v$ and $V$ are the same. Then $\varphi(z) = \psi(\mu(z-1))$ and $\varphi''(1) = \psi''(0)\mu^2$. Combining this equality with (A.38) we have

$$\psi''(0) = \psi''(0)\Lambda_2\mu^{-2}(1-\Lambda_1)^{-3} + 2, \tag{A.39}$$

By definition, all terms in this equality are nonnegative, $\Lambda_2\mu^{-2} \geq 1$, and $(1-\Lambda_1)^{-3} > 1$. Therefore (A.39) is contradictory.